\documentclass{article}
\usepackage{ijcai25}

\usepackage{times}
\usepackage{soul}
\usepackage{url}
\usepackage[hidelinks]{hyperref}
\usepackage[utf8]{inputenc}
\usepackage[small]{caption}
\usepackage{graphicx}
\usepackage{amsmath}
\usepackage{amsthm}
\usepackage{booktabs}
\usepackage{algorithm}
\usepackage{algorithmic}
\usepackage[switch]{lineno}
\usepackage{amssymb}
\usepackage{xcolor}
\usepackage{multirow}

\title{AniSora: Exploring the Frontiers of Animation Video Generation in the Sora Era}

\author{
Yudong Jiang \textsuperscript{*}\textsuperscript{†} 
\and Baohan Xu \textsuperscript{*}\textsuperscript{†}
\and Siqian Yang\textsuperscript{*}\textsuperscript{†}
\and Mingyu Yin\textsuperscript{†} \and Jing Liu\textsuperscript{†}
\and Chao Xu \and Siqi Wang \and Yidi Wu \and Bingwen Zhu \and Yue Zhang
\and Jinlong Hou \and Huyang Sun
\affiliations
Bilibili Inc.
}

\begin{document}

\maketitle
\begin{figure*}[h]
    \centering
    \includegraphics[scale=0.248]{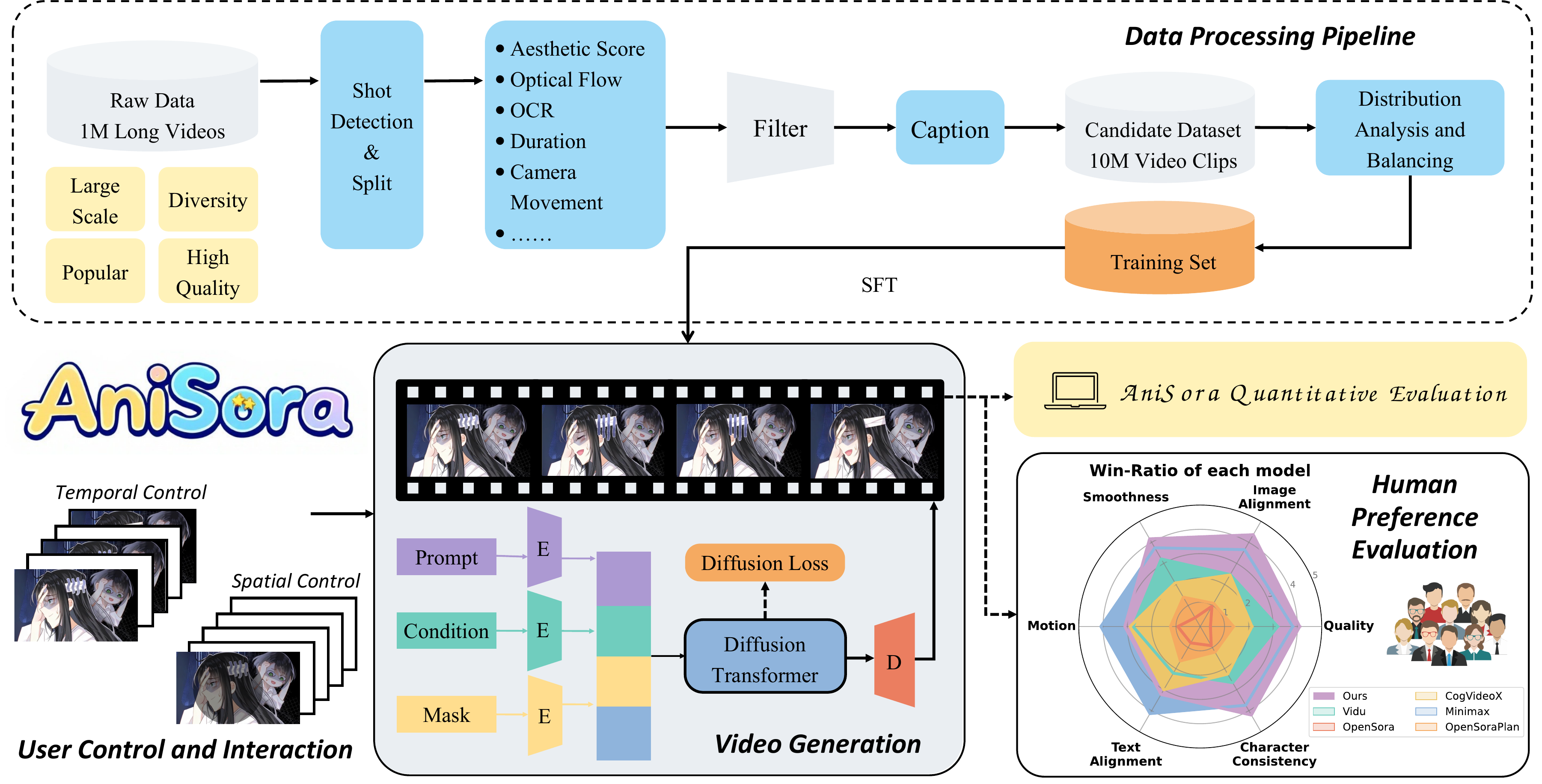}
    \caption{\textbf{Overview.} We propose  \textbf{AniSora}, a comprehensive framework for animation video generation that integrates a high-quality animation dataset, a spatiotemporal conditional model, and a specialized animation video benchmark. The \textbf{Data Processing Pipeline} constructs a 10M video clip dataset derived from 1M diverse long animation videos. The \textbf{Video Generation} model employs a spatiotemporal conditional model, supporting various \textbf{User Control and Interaction} modes and enabling tasks such as frame interpolation, localized guidance, and so on. The benchmark set comprises 948 ground-truth videos spanning diverse styles, common motions, and both 2D and 3D animations. The prompt suite provides standardized prompts and guiding conditions, complemented by a \textbf{Quantitative Evaluation} with six objective metrics for assessing visual appearance and consistency. Additionally,  \textbf{Human Preference Evaluation} confirms strong alignment with the proposed metrics. \textbf{AniSora} surpasses SOTA models, establishing a new benchmark for animation video generation.}
    \label{fig:cover}
\end{figure*}

\begin{abstract}
Animation has gained significant interest in the recent film and TV industry. Despite the success of advanced video generation models like Sora, Kling, and CogVideoX in generating natural videos, they lack the same effectiveness in handling animation videos. Evaluating animation video generation is also a great challenge due to its unique artist styles, violating the laws of physics and exaggerated motions. In this paper, we present a comprehensive system, \textbf{AniSora}, designed for animation video generation, which includes a data processing pipeline, a controllable generation model, and an evaluation benchmark. Supported by the data processing pipeline with over 10M high-quality data, the generation model incorporates a spatiotemporal mask module to facilitate key animation production functions such as image-to-video generation, frame interpolation, and localized image-guided animation. We also collect an evaluation benchmark of 948 various animation videos, with specifically developed metrics for animation video generation.
   \textbf{Our entire project is publicly available on  }\href{https://github.com/bilibili/Index-anisora/tree/main}{\textcolor{red}{\textit{https://github.com/bilibili/Index-anisora/tree/main}}}
\end{abstract}

\section{Introduction}
\label{sec:intro}

\footnotetext{\textsuperscript{*}Equal contributions. \textsuperscript{†}Core contributors: \{jiangyudong, xubaohan, yangsiqian, yinmingyu, liujing\}@bilibili.com}
The animation industry has seen significant growth in recent years, expanding its influence across entertainment, education, and even marketing. As demand for animation content rises, the need for efficient production processes is also growing quickly, particularly in animation workflows. Traditionally, creating high-quality animation has required extensive manual effort for tasks like creating storyboards, generating keyframes, and inbetweening, making the process labor-intensive and time-consuming. Previous efforts\cite{siyao2021anime,xing2024tooncrafter} to incorporate computer vision techniques have assisted animators in generating inbetween frames for animation. However, these methods often show effectiveness only within certain artistic styles, limiting their applicability to the varied demands of modern animations.

With recent advancements in video generation, there has been notable progress in generating high-quality videos across various domains. Inspired by Generative Adversarial Networks\cite{goodfellow2014generative}, Variational Autoencoders\cite{kingma2013auto}, and, more recently, transformer-based architectures\cite{vaswani2017attention,peebles2023scalable}, the field has seen remarkable improvements in both efficiency and output quality. However, most video generation methods are trained and evaluated on general-purpose datasets, typically featuring natural scenes or real-world objects\cite{blattmann2023stable,yang2024cogvideox}. The domain of animation video generation, which plays an important role ranging from entertainment to education, has received relatively little attention. Animation videos often rely on non-photorealistic elements, exaggerated expressions, and non-realistic motion, presenting unique challenges that current methods do not address.

In addition to the generation challenges, the evaluation of video synthesis is also inherently complex. Evaluating video generation quality requires assessing not only the visual fidelity of each frame but also temporal consistency, coherence, and smoothness across frames\cite{huang2024vbench}. For animation video generation, this challenge is amplified. Animation videos feature unique artist styles, including color and style that need to be consistent, even as characters undergo exaggerated motions and transformations. Traditional evaluation metrics are commonly used for real-world videos, which may not fully capture the consistency of the main characters and art style of this kind of video. Therefore, developing effective evaluation datasets and metrics customized for animation video generation is essential in this specialized field.

\begin{figure}[!h]
    \centering
    \includegraphics[scale=0.35]{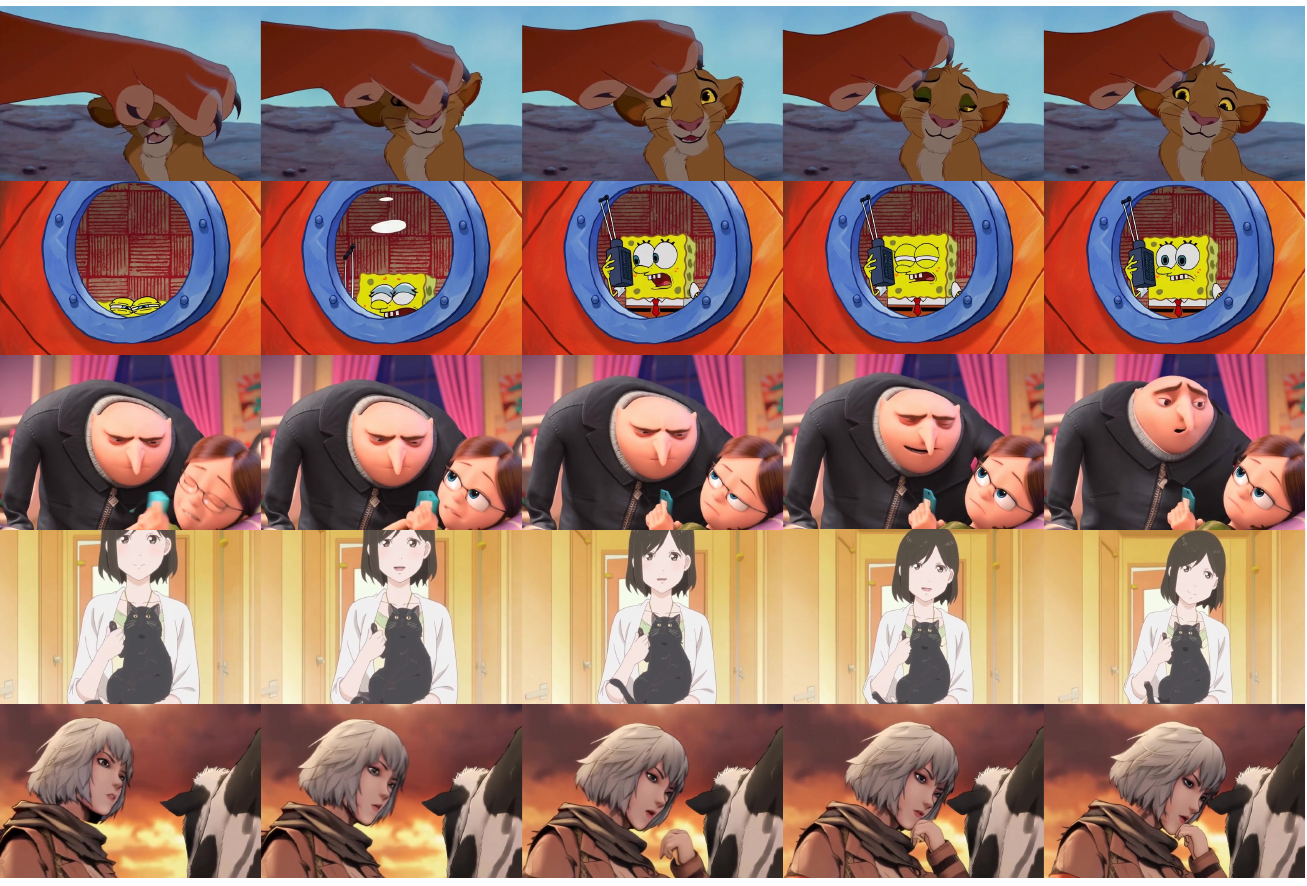}
    \caption{Our method can generate high quality and high consistency in various kinds of 2D/3D animation videos. These examples are generated under image-to-video settings conditioned on the leftmost frame. It is best viewed in color.}
    \label{fig:showcase}
\end{figure}

In this paper, as shown in Fig. \ref{fig:cover}, a full system \textbf{AniSora} is presented for animation video generation. First, our data processing pipeline offers over 10 million high-quality text-video pairs, forming the foundation of our work. Secondly, we develop a unified diffusion framework adapted for animation video generation. Our framework leverages spatiotemporal masking to support a range of tasks, including image-to-video generation, keyframe interpolation, and localized image-guided animation. By integrating these functions, our system bridges the gap between keyframes to create smooth transitions and enables dynamic control over specific regions, such as animating different characters speaking precisely. This allows a more efficient creative process for both professional and amateur animation creators. Fig. \ref{fig:showcase} demonstrates some examples generated by our model under image-to-video conditions.

Additionally, we propose a benchmark dataset and evaluation metrics specifically designed for animation video evaluation. Currently, there is no existing evaluation dataset for this purpose, so we collected 948 animation videos across various categories and manually refined the prompts. Besides, existing evaluation standards struggle to effectively assess the quality of animation video generation. To address this gap, we have introduced innovative metrics that specifically assess animation video generation. These include character consistency, animation art style consistency, and distortion detection,  which are crucial for maintaining the unique visual identity of animation videos.

Our contributions can be summarized as follows:
\begin{itemize}
    \item{We develop a comprehensive video processing system that significantly enhances preprocessing for animation video generation.}
    \item{We propose a unified framework designed for animation video generation with a spatiotemporal mask module, enabling tasks such as image-to-video generation, frame interpolation, and localized image-guided animation.}
    \item{To the best of our knowledge, we for the first time released a benchmark dataset and evaluation metrics specifically for evaluating animation video generation.}
\end{itemize}

\section{Related Work}

\subsection{Video generation models}
    With the development of diffusion models, significant progress has been made in video generation over the past two years. Some research including\cite{blattmann2023stable,yang2024cogvideox,opensora,pku_yuan_lab_and_tuzhan_ai_etc_2024_10948109} have demonstrated promising results in general video generation. Due to the limited available animation datasets, these models are not particularly effective for animation video generation.

\subsection{Animation video datasets}
    Video data is one of the most critical elements for generation models, particularly for domain-specific data. However, obtaining high-quality animation video data is especially difficult compared to natural video datasets. Previous research has released some animation-related datasets, including ATD-12K\cite{siyao2021anime}, AVC\cite{wu2022animesr}. While these datasets, collected from various animation movies, are helpful for interpolation and super-resolution tasks, they are limited by small size. More recently, Sakuga-42M\cite{sakuga42m2024} has been proposed with 1.2M clips. It has improved compared to previous datasets that only contained a few hundred clips. However, this remains insufficient for training video generation models, in contrast to general video data sets such as Panda-70M\cite{chen2024panda} and InternVid-200M\cite{wang2023internvid}. Additionally, 80\% of its clips are low-resolution and less than 2 seconds, which hampers the generation of high-quality videos.

\subsection{Evaluation of video generation models}
    Evaluating video generation models has remained a significant challenge in the past few years. Recently, Liu et al. have made great efforts to generate a diverse and comprehensive list of 700 prompts using LLM\cite{liu2023evalcrafter}. Besides, Huang et al. have proposed vbench for general video generation\cite{huang2024vbench}. The authors have released 16 evaluation dimensions and prompt suites. While these dimensions are still insufficient to comprehensively evaluate all aspects of animation video generation. Moreover, there is a notable absence of dedicated animation evaluation datasets, which limits the ability to benchmark models specifically designed for this genre. In \cite{zeng2024dawn}, the authors have focused mainly on the performance of recent video generation models in various categories of datasets. Furthermore, they have also investigated some vertical-domain models like pose-controllable generation and audio-driven animation.
    
    While these works provide valuable insights into the capabilities of these models in generating diverse video content, they don't specifically address the unique requirements and challenges associated with animation video generation. 

    

\section{Dataset}
\label{sec:dataset}
We build our animation dataset according to the observation that \emph{high quality text-video pairs are the cornerstone of video generation}, which is proved by recent researches \cite{polyak2024movie}.
In this section, we give a detailed description of our animation dataset and the evaluation benchmark.

\noindent\textbf{Animation Dataset Construction:}
We build a pipeline to get high-quality text-video pairs among $1$ million raw animation videos.
First of all, we use scene detection\cite{pyscenedetect} to divide raw videos into clips.
Then, for each video clip, we construct a filter rule from four dimensions: text-cover region, optical flow score, aesthetic score, and number of frames. 
The filter rule is gradually built up through the observations in model training.
In detail, the text-cover region score\cite{baek2019character} can drop those clips with text overlay similar to end credits.
Optical flow score \cite{raft} prevents those clips with still images or quick flashback scenes. 
Aesthetic score \cite{aesthetic} is utilized to preserve clips with high artistic quality.
Besides, we retain the video clips whose duration is among $2s$-$20s$ according to the number of frames.
\textcolor{black}{Furthermore, we collected 0.5M high-quality animation videos along with their corresponding captions to create video-text pairs, which were used to fine-tune Qwen-VL2\cite{Qwen2VL}. After fine-tuning, the model provides more accurate descriptions of characters, scenes, and action details in animation content.} After the steps mentioned above, about $10\%$ clips (more than 10M clips) with captions can be retained in the training step.

\textcolor{black}{In addition, since occupationally-generated animation videos typically have significantly higher production costs and quality compared to user-generated animation content, we fine-tuned Qwen-VL2 based on these data. This model is then utilized to filter higher-quality clips to further improve the model's performance.}
Specifically, during the training process, we adjust the proportions of specific training data (e.g., talking and motion amplitude) according to the observed performance.




\noindent\textbf{Benchmark Dataset Construction:}
Since there is currently no benchmark dataset specifically designed for animation content, we construct a benchmark dataset manually to compare the generation videos between our model and other recent researches.
$948$ animation clips are collected and labeled with different actions, e.g., talking, walking, running, eating, and so on. 
Among them, there are $857$ 2D animation clips and $91$ 3D clips.
These action labels are summarized from more than $100$ common actions with human annotation.
Each label contains $10$-$30$ video clips. 
\textcolor{black}{The corresponding text prompt is generated by fine-tuned Qwen-VL2 (mentioned above) at first, then is corrected manually to guarantee the text-video alignment. (More details in section \ref{sec:benchmark} and \ref{sec:experiment})}


\section{Method}
In this section, we present an effective approach for animation video generation using a diffusion transformer architecture. Section \ref{sec:Dit-based Video Generation Model} provides an overview of the foundational diffusion transformer model. In section \ref{sec:Spatiotemporal Condition Model}, we introduce a spatiotemporal mask module that extends the model, enabling crucial animation production functions such as image-to-video generation, frame interpolation, and localized image-guided animation within a unified framework. These enhancements are essential for professional animation production. Finally, section \ref{sec:Supervised Fine-Tuning} details the supervised fine-tuning strategy employed on the animation dataset.

\subsection{DIT-based Video Generation Model}
\label{sec:Dit-based Video Generation Model}
\begin{figure}
    \centering
    \includegraphics[width=0.95\columnwidth]{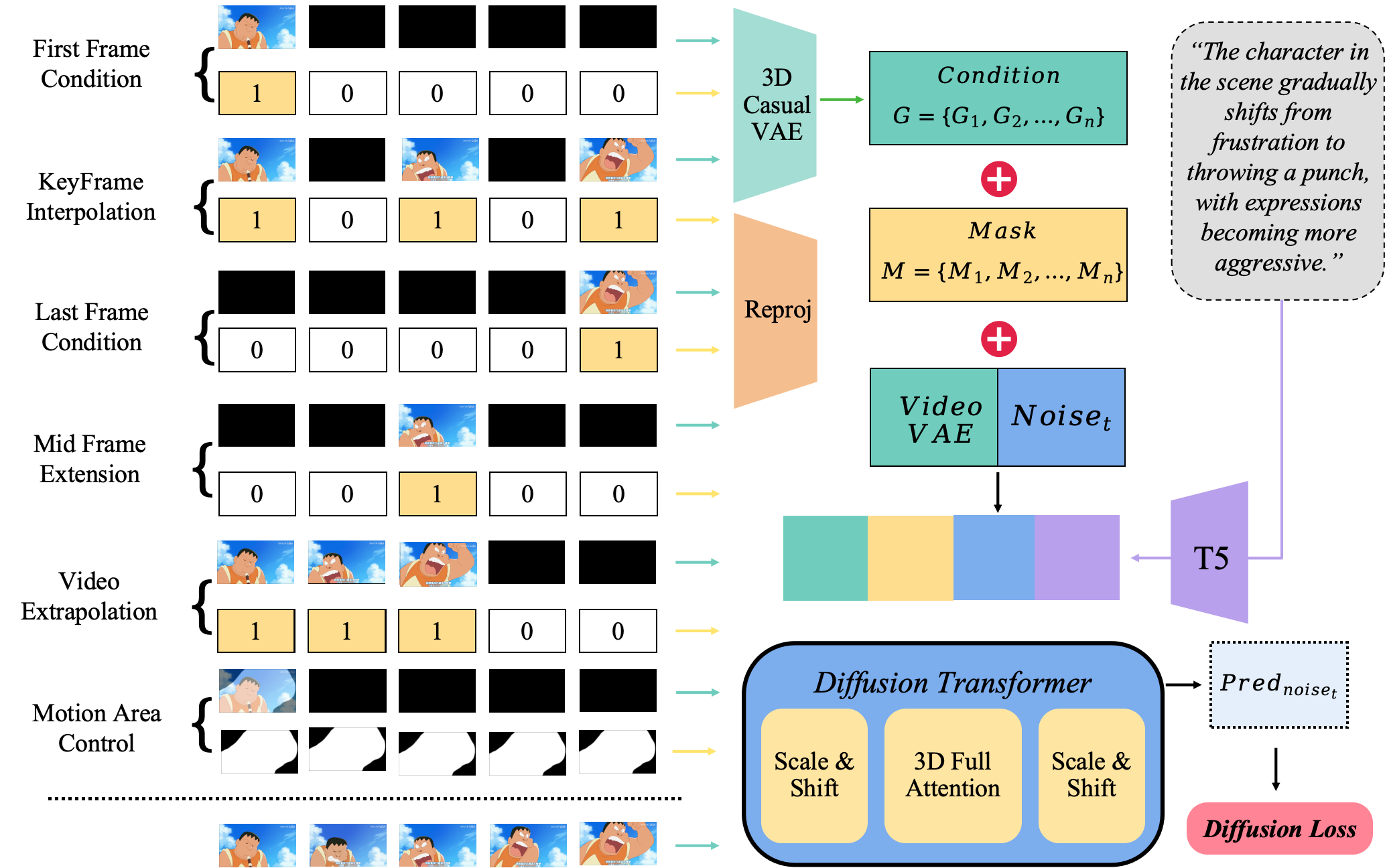}
    \caption{\textbf{Method.} This figure illustrates the Masked Diffusion Transformer framework for animation video generation, designed to support various spatiotemporal conditioning methods for precise and flexible animation control. A 3D Causal VAE compresses spatial-temporal features into a latent representation, generating the guide feature sequence $G$, while a reprojection network constructs the mask sequence $M$. These components, combined with noise and prompt's feature, serve as input to the Diffusion Transformer. The transformer employs techniques such as patchify, 3D-RoPE embeddings, and 3D full attention to effectively capture spatial-temporal dependencies. This framework seamlessly integrates keyframe interpolation, motion control, and mid-frame extension, simplifying animation production and enhancing creative possibilities.}
    \label{fig:framework}
\end{figure}

We adopt a  DiT-based\cite{peebles2023scalable}   text-to-video diffusion model as the foundation model. As shown in Fig. \ref{fig:framework}, the model leverages the three components to achieve coherent, high-resolution videos aligned with text prompts. 

\noindent\textbf{3D Casual VAE} used in video generation frameworks\cite{gupta2023photorealistic,yu2023language}serves as a specialized encoder-decoder architecture tailored for spatiotemporal data compression. This 3D VAE compresses videos across both spatial and temporal dimensions, significantly reducing the diffusion model computing. We follow the approach of Yang et al. \cite{yang2024cogvideox} to extract latent features, transforming the original video with dimensions \( (W, H, T, 3) \) into a latent representation of shape \( (W/8 ,H/8 ,T/4 ,16) \).

\noindent\textbf{Patchify} is a critical step for adapting vision tasks to transformer-based architectures \cite{dosovitskiy2020image}. Given an input video of size \( T \times H \times W \times C \), it is split spatio into patches of size \( P \times P \), and temporal into size \(\ Q \) resulting in \( (T/Q) \times (H/P) \times (W/P) \times C \) patches. This method enables efficient high-dimensional data processing by reducing complexity while retaining local spatial information. 

\noindent\textbf{3D Full Attention} We propose 3D full attention module for spatial and temporal modeling considering the great success of long-context training in LLM\cite{dubey2024llama} and foundation video generation model\cite{yang2024cogvideox,polyak2024movie}. 

\noindent\textbf{Diffusion schedule} applies Gaussian noise to an initial sample \( x_0 \) over \( T \) steps, generating noisy samples \( x_t = \sqrt{\alpha_t} \, x_0 + \sqrt{1 - \alpha_t} \, \epsilon \), where \( \alpha_t = \prod_{i=1}^t (1 - \beta_i) \) and \( \epsilon \sim \mathcal{N}(0, I) \). The reverse process predicts \( \epsilon \) by minimizing the mean squared error: 
\[
\mathcal{L}_{\text{diffusion}} = \mathbb{E}_{x_0, \epsilon, t} \left[ \| \epsilon - \epsilon_\theta(x_t, t) \|_2^2 \right].
\] 
To stabilize training, we use the v-prediction loss \cite{salimans2022progressive}, where \( v = \sqrt{1 - \alpha_t} \, x_0 - \sqrt{\alpha_t} \, \epsilon \) and the loss becomes 
\[
\mathcal{L}_{v-prediction} = \mathbb{E}_{x_0, v, t} \left[ \| v - v_\theta(x_t, t) \|_2^2 \right].
\] 
This approach enhances stability and model performance.
\subsection{Spatiotemporal Condition Model}  \label{sec:Spatiotemporal Condition Model}
\noindent\textbf{Keyframe Interpolation} creates smooth transitions between key-frames by generating intermediate frames, or "in-between." It is an essential stage in professional animation production and represents some of the most labor-intensive tasks for artists. We extend this concept to video generation conditioned on one or multiple arbitrary frames placed at any position within a video sequence. 

\noindent\textbf{Motion Control} Our framework enables precise control over motion regions addressing the limitations of text-based control in these aspects. This approach enhances artists' control over video content, allowing them to express their creativity while significantly reducing their workload.

\subsubsection{Masked Diffusion Transformer Model}
\label{sec:Masked Diffusion Transformer Model}
In the Masked Diffusion Transformer framework, we construct a guide feature sequence \( G = \{G_1, G_2, \dots, G_n\} \) by placing the VAE-encoded guide frame \( F_{p_i} \) at designated positions \( p_i \), while setting \( G_j = 0 \) for all other positions \( j \neq p_i \). A corresponding mask sequence \( M = \{M_1, M_2, \dots, M_n\} \) is generated, where \( M_{p_i} = 1 \) for guide frame positions and \( M_j = 0 \) otherwise. The mask is processed through a re-projection function, yielding an encoded representation \( Reproj(M) \). The final input to the Diffusion Transformer is the concatenation of noise, encoded mask, prompt's T5 feature, and guide sequence along the channel dimension:
\begin{equation}
X = Concat (Noise_{t}, Reproj(M), G, T5)
\label{eq:maskedDiffusion}
\end{equation}
This setup integrates position-specific guidance and mask encoding, enhancing the model’s conditioned generation capabilities.

\subsubsection{Motion Area Condition}
\label{sec:Motion Area Condition}
This framework can also support spatial motion area conditions inspired by Dai et.al\cite{dai2023animateanything}. Given the image condition \( F_{p_i} \), and motion area condition is represented by mask \(M_{F}\), the same shape with  \( F_{p_i} \). Motion area in \(M_{F}\) is labeled 1, other place is set to 0. As equation \ref{eq:maskedDiffusion} in \ref{sec:Masked Diffusion Transformer Model}, for guide frame position \( p_i \), set \( M_{p_i} = M_{F} \).
The data processing and training pipeline can be summarized as follows:
\textbf{Constructing video-mask pairs,} we first construct paired training data consisting of videos and corresponding masks. Using a foreground detector by Kim et.al \cite{kim2022revisiting}, we detect the foreground region in the first frame of the video. This region is then tracked across subsequent frames to generate a foreground mask for each frame.
\textbf{Union of foreground masks,} the per-frame foreground masks are combined to create a unified mask \(M_{F}\), representing the union of all foreground regions across the video.
\textbf{Video latent post-processing,} for the video latent representation \(z_{0}\), non-moving regions are set to the latent features of the guide image, ensuring static areas adhere to the guide.
\textbf{LoRA-based conditional training,} we train the conditional guidance model using Low-Rank Adaptation (LoRA) with a parameter size of 0.27B. This approach significantly reduces computational requirements while enabling efficient model training.
\subsection{Supervised Fine-Tuning}
\label{sec:Supervised Fine-Tuning}
We initialize our model with the pre-trained weights of CogVideoX, which was trained on 35 million diverse video clips. Subsequently, we perform full-parameter supervised fine-tuning (SFT) on a custom animation training dataset to adapt the model specifically for animation tasks.

\noindent\textbf{Multi-Task Learning} Compared to the physically consistent motion patterns in the real world, animation styles, and motion dynamics can vary significantly across different works. This domain gap between datasets often leads to substantial quality differences in videos generated from guide frames with different artistic styles. We incorporate image generation into a multi-task training framework to improve the model's generalization across diverse art styles. Experimental results in the appendix demonstrate that this approach effectively reduces the quality gap in video generation caused by stylistic differences in guide frames.

\noindent\textbf{Mask Strategy} During training, we unmask the first, last, and other frames obtained through uniform sampling with a 50\%  probability. This strategy equips the model with the ability to handle arbitrary guidance, enabling it to perform tasks such as in-betweening, first-frame continuation, and arbitrary frame guidance, as discussed in Section \ref{sec:Masked Diffusion Transformer Model}.

\textcolor{black}{In practice, we also employed several other effective training strategies, such as \textit{weak to strong training}, \textit{generated subtitle removal}, and \textit{temporal multi-resolution training}. Detailed training procedures can be found in the appendix.}

\section{Benchmark}
\label{sec:benchmark}
To evaluate the effects of the animation video generation models, we build a comprehensive benchmark dataset, as mentioned in section \ref{sec:dataset}.
To give a fair comparison of different methods, we define $6$ basic dimensions to describe the quality of the generated videos. 
Then, we introduce the human annotation and evaluation metrics, which are partly based on the annotation results.

\subsection{Evaluation Dimensions}
\label{subsec:evaluation}
In fact, the essential concepts for evaluating a animation video generation model are \textit{visual appearance} and \textit{visual consistency}.
Visual appearance describes the basic quality, which is only concerned with the generation of videos themselves, including visual smoothness, visual motion, and visual appeal, while visual consistency considers the text-video, image-video, and character consistency, respectively.

\begin{figure}
    \centering
    \includegraphics[scale=0.26]{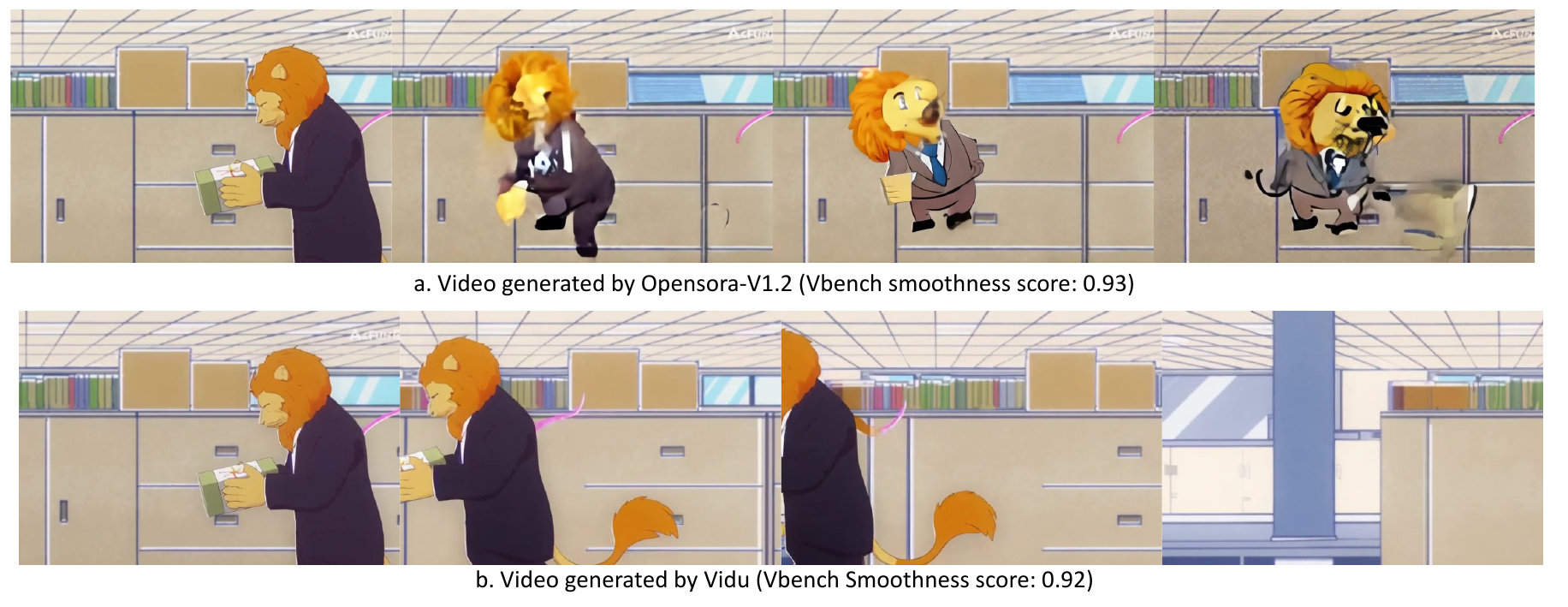}
    \caption{Video generated by Opensora-V1.2 (top) and Vidu (bottom). The top video received a higher score despite containing noticeable distortions.}
    \label{fig:smoothness}
\end{figure}

Inspired by Vbench\cite{huang2024vbench}, we first adopted a similar approach to evaluate visual appearance and consistency, such as calculating the clip score between frames to analyze distortions and using aesthetic scores to estimate aesthetic quality. However, we found that these scores often showed significant discrepancies from subjective human experiences on animation data and there was no clear distinction between different methods. As shown in Fig.\ref{fig:smoothness}, the bottom video received a lower score due to its larger motion amplitude. However, the top video contains visible distortions that are easily noticeable to humans, yet these are not reflected in the scoring metrics. 

Therefore, we adopted a regression-based approach for certain dimensions to learn human scoring standards. A detailed description of six evaluated metrics is given as follows.
These evaluation criteria are specifically designed for animation data and align closely with human subjective experience.

\subsection{Human Annotation}\label{sec:human_annotation}
In the experiment, each case in the benchmark has $6$ generation video clips with different participant models. $20$ expert volunteers give their rating (1-5, 5 is the best) from above $6$ dimensions, the detail scores are shown in Section \ref{subsec:benchmark results}.

\subsection{Visual Appearance} \label{sec:visual_appearance}
We evaluate the basic quality of a video clip from three aspects: visual smoothness, visual motion, and visual appeal.
\textcolor{black}{We constructed an evaluation model training set to develop an evaluation model aligned with human scoring. 
We generated $5000$ animated video clips using various models described in Section \ref{subsec:benchmark results}. These samples were manually annotated following the approach outlined in Section \ref{sec:human_annotation}, serving as the ground truth for training the evaluation model.}


\noindent\textbf{Visual Smoothness}
Our goal was to learn the human standards to evaluate video smoothness and to be able to identify animated videos with distortions. 
Thus, we trained a model to regress human-generated scores to avoid the impact of motion on visual content.
\textcolor{black}{The training set consists of generated clips and annotations, mentioned in Section \ref{sec:visual_appearance}.}
To enhance the robustness of the model, we also incorporated hundreds of anime videos as the highest-scoring examples.




\noindent\textbf{Visual Motion} We employ a model based on the ActionCLIP \cite{wang2021actionclip} framework to train a motion-scoring model that evaluates the magnitude of the primary motion in animation videos. 
About $2$ thousand animation video clips and their corresponding motion captions are collected into $6$ degrees of movement amplitude (from stillness to significant motion) to finetune the action model.
Finally, the motion score is obtained from the similarity score between the designed motion prompt and the participant video.
\begin{equation}
    S_{motion} = Cos(MCLIP(V), MCLIP(T_{m})),
\end{equation}
where $MCLIP$ denotes the finetuning action model. $V$ represents the generation video and $T_{m}$ denotes the designed motion prompt. (Details is provided in supplementary)

\noindent\textbf{Visual Appeal}
We define visual appeal score to reflect the basic effects of video generation. As discussed in Section \ref{subsec:evaluation}, the aesthetic score models used in previous studies were trained on real-world datasets such as LAION-5B, simply calculating aesthetic scores on animation data does not provide sufficient differentiation, as the results from various methods are indistinguishable.

Using the key frame extraction method to collect the key frames in the video first, then we train a regress model on the evaluation model training set to learn human aesthetic standards.
The formulation shows as follows:
\begin{equation}
    S_{appeal}=Aes(SigLIP(I_{0,1,\dots, K})) \quad I_{i}\in{KeyFrm(V)},
\end{equation}
where $KeyFrm$, $SigLIP$ and $Aes$ denote the key frame extraction method, feature encoder method and aesthetic evaluation method, and $K$ denotes the number of the keyframes.

\subsection{Visual Consistency}
Three factors are considered to evaluate the visual consistency of the generation video: text-video, image-video, and character consistency, respectively.

\noindent\textbf{Text-video Consistency}
\textcolor{black}{To evaluate the text-video consistency, we finetune the vision encoder and the text encoder modules with a regression head according to animation video-text pairs according to the training set in Section \ref{sec:visual_appearance}.} 
The formulation is shown as follows:
\begin{equation}
    S_{tvc}=Reg(E_{v}, E_{t}),
\end{equation}
where $Reg$ denotes the regression head, and $E_{v}$, $E_{t}$ denote the vision and text encoder.

\noindent\textbf{Image-video Consistency}
In the I2V situation, the participant image, as an input factor, should ensure that its style is consistent with the generated videos.
Similar to text-video consistency, we combine a vision encoder with a regression head to evaluate the score. \textcolor{black}{The model is also fine-tuned on the training set in Section \ref{sec:visual_appearance}.}
The formulation lists as follows:
\begin{equation}
    S_{ivc}=Reg(E_{v}(V), E_{v}(I_{p})),
\end{equation}
where $V$ and $I_{p}$ denote the participant video clip and the input image.


\noindent\textbf{Character Consistency}
Character consistency is a crucial factor in animation video generation. 
When the character generated by the protagonist in the animation changes, even if the quality of the video is great, it still has the risk of infringement.
Hence, we design a set of procedures including detection, segmentation, and recognition. 
We apply GroudingDino \cite{ren2024grounding} and SAM \cite{ravi2024sam2segmentimages} to achieve character mask extraction for each frame in the videos. 
Then, we finetune a BLIP-based model \cite{li2022blip} to establish connections between each mask and the animation IP character. 
In detail, thousands of source video clips with their characters labeled are treated as training sets to obtain and store the reliable features from BLIP-based model.
In the evaluation step, we get the score of character consistency by calculating the cosine similarity between the generated and stored character's features.
\begin{equation}
    S_{IPc}=\frac{1}{S}\sum^{S}_{i}Cos(BLIP(M_{i}), fea_{c}),
\end{equation}
where $S$ denotes the number of sample frames, $M_{i}$ denotes the mask obtained from GroudingDino and SAM methods, and $fea_{c}$ denotes the stored character's features.




\section{Experiment}
\label{sec:experiment}





\begin{table*}[ht]
    \centering
    \caption{Benchmark Evaluation Results}
    \label{tab:benchmark_results}
    \begin{tabular}{lccccccc}
        \toprule
        Models & Human & Visual & Visual & Visual & Text-Video & Image-Video & Character \\
        & Evaluation & Smooth & Motion & Appeal & Consistency & Consistency & Consistency \\
        \midrule
        Vidu-1.5 & 60.98 & 55.37 & \textbf{78.95} & 50.68 & 60.71 & 66.85 & 82.57 \\
        Opensora-V1.2  & 41.10 & 22.28 & 74.9 & 22.62 & 52.19 & 55.67 & 74.76 \\
        Opensora-Plan-V1.3  & 46.14 & 35.08 & 77.47 & 36.14 & 56.19 & 59.42 & 81.19 \\
        CogVideoX-5B-V1  & 53.29 & 39.91 & 73.07 & 39.59 & 67.98 & 65.49 & 83.07 \\
        MiniMax-I2V01 & 69.63 & 69.38 & 68.05 & \textbf{70.34} & \textbf{76.14} & 78.74 & 89.47 \\
        \midrule
        \textbf{AniSora(Ours)} & \textbf{70.13} & \textbf{71.47} & 47.94 & 64.44 & 72.92 & \textbf{81.54} & \textbf{94.54} \\
        AniSora(Interpolated Average) & - & 70.78 & 53.02 & 64.41 & 73.56 & 80.62 & 91.59 \\
        AniSora(KeyFrame Interpolation) & - & 70.03 & 58.1 & 64.57 & 74.57 & 80.78 & 91.98 \\
        \bottomrule
    \end{tabular}
\end{table*}

\begin{figure*}
    \centering
    \includegraphics[scale=0.26]{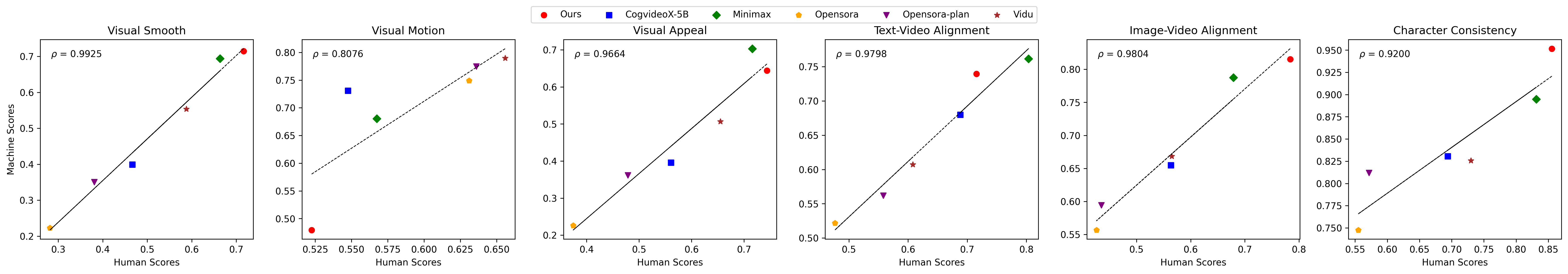}
    \caption{Human Evaluation and Benchmark Results Alignment}
    \label{fig:human_bench_alignment}
\end{figure*}

\subsection{Benchmark Evaluation}
\label{subsec:benchmark results}

In this section, we give both objective and human evaluation results of our benchmark.
Six recent i2v models are involved in our evaluation: Open-sora \cite{opensora}, Open-sora-plan \cite{pku_yuan_lab_and_tuzhan_ai_etc_2024_10948109}, Cogvideox \cite{yang2024cogvideox}, Vidu-1.5 \cite{vidu}, Minimax-I2V01 \cite{minimax} and Anisora(ours).
Tab. \ref{tab:benchmark_results} gives the detailed scores from $6$ dimensions in the benchmark evaluation and the overall scores of the human evaluation. 
We observe that our model performs better than the other five methods on most dimensions, especially on visual smoothness and character consistency, except on the visual motion dimension.
These mainly because we conduct a thorough assessment of the balance between generation quality and motion magnitude, and find most generation clips with big motion results in distortion or unnatural segments.
\textcolor{black}{
In order to prove our benchmark is applicable to anime scenarios, we also evaluated our benchmark dataset using seven relevant dimensions in VBench benchmark. 
Due to the space limitation, we outline the results in Appendix. 
We observed that certain dimensions, including Motion Smoothness, Aesthetic Quality, I2V Background, and Overall Consistency, lacked sufficient discriminative power. In particular, some poorly generated results received higher scores than the ground truth, highlighting a discrepancy that fails to accurately capture human perception and experience.}

Fig. \ref{fig:human_bench_alignment} illustrates the detailed correlations among $6$ dimensions between human evaluation and benchmark results.
Obviously, they are highly correlated with each other.



\subsection{Spatiotemporal Mask Module}
\noindent\textbf{Frame Interpolation}
Tab. \ref{tab:benchmark_results} presents the results of different interpolation settings on benchmark dataset. Our evaluation process involved generating videos on our benchmark with various guidance conditions sampled at equal proportions, which can refer to Fig .\ref{fig:framework}. We then compute the average score of all samples as well as a specific statistical analysis for keyframe interpolation results. The performance indicates that single-frame guidance achieves competitive results whether the guiding frame is placed at the beginning, middle, or end of the frame sequence, which also consistently outperforms other methods. 
Adding more guiding frames further improves character consistency. We also observed from the motion and smooth score that our baseline model achieves a balance between motion range and consistency, while keyframe guidance enables the model to produce animation videos with larger motion ranges and more realistic motion. More samples can be found in the appendix. 



\noindent\textbf{Motion Area Condition}
The evaluation of motion area condition is constructed based on our benchmark dataset. For each initial frame, we performed saliency segmentation, followed by connected-component analysis to generate bounding boxes for each instance. Then we manually filtered the results to select high-quality motion area masks, resulting in 200 samples. Following the experiment settings in \cite{dai2023animateanything}, we 
conducted the comparison of motion mask precision in Tab. \ref{tab:motion_mask_precision}. We also computed the score of AnimateAnything on our selected 200 samples. The lower score is primarily due to flickering and noise appearing outside the motion mask area. The results demonstrate the effectiveness of our spatial mask module in controlling movable regions. It is also noticeable that even without motion control, our generation model trained for animation video still shows a certain level of control. This may due to the effective prompt-based guidance for the main subject. Motion mask guidance examples are shown in the appendix.

\begin{table}[h]
    \centering
    \caption{Comparison of motion mask precision}
    \begin{tabular}{l|c}
        \midrule
         \small{\textbf{Method}} &  \small{\textbf{Motion Mask Precision}} \\
        \midrule
        \small{AnimateAnything} &  \small{0.6141} \\
         \small{Ours - No Control} &  \small{0.4989} \\
         \small{Ours - Motion Mask} &  \small{\textbf{0.9604}} \\
        \midrule
    \end{tabular}
    \label{tab:motion_mask_precision}
\end{table}


\subsection{Animation Video Training}
\noindent\textbf{2D and 3D Animation} 
Analysis using QWEN2 \cite{Qwen2VL} shows that 2D samples account for 85\% of our data set, yet the quality of 3D animation generation consistently exceeds that of 2D. Benchmark evaluations in the appendix confirm 3D animations demonstrate superior visual appearance and consistency, a phenomenon unique to animation training. We attribute this gap to the pre-trained model's exposure to real-world video data. Unlike 2D animations with diverse motion patterns, 3D animations rendered by physics-based engines like Unreal Engine follow consistent physical laws, enabling better knowledge transfer during SFT. Consequently, improving generalization on 2D animation data remains more challenging than on 3D or real-world data.

\noindent\textbf{Multi-Task Learning} 
We evaluated multi-task training using a manga with a unique artistic style. About 270 illustrations were used for the image generation task, while video training data remained the same as the baseline model. Additional illustrations served as first-frame conditions during video generation. After 5k training steps, as shown in the appendix, the generated videos showed significantly greater stability and improved visual quality, particularly with highly distinctive guidance images. This approach effectively tailors animations to specific characters and mitigates domain gaps caused by variations in artistic styles, especially when high-quality animation data is limited.

\noindent\textbf{Low-resolution vs High-resolution}
During the weak-to-strong training process, we observed that higher frame rates and resolutions enhance stability in visual details.  As demonstrated in the appendix, at 480P, facial features exhibit noticeable distortions, while at 720P, the model preserves both motion consistency and fine details. The higher resolution increases token representation for high-density areas, improving temporal consistency and overall content quality.

\section{Conclusion}
In this paper, our proposed \textbf{AniSora}, a unified framework provides a solution to overcoming the challenges in animation video generation. Our data processing pipeline generates over 10M high-quality training clips, providing a solid base for our model. Leveraging a spatiotemporal mask, the generation model can create videos based on diverse control conditions. Furthermore, our evaluation benchmark demonstrates the effectiveness of our method in terms of character consistency and motion smoothness. We hope that our research and evaluation dataset establish a new benchmark and inspire further work in the animation industry. 
Besides, we are going to evaluate more models on our benchmark, providing valuable insights for model optimization.

Despite promising results, some artifacts and flickering are still present in the generated videos. In future work, we plan to integrate reinforcement learning with our evaluation benchmark to generate higher-quality videos.

\section*{Acknowledgments}

We sincerely thank Qiang Zhang, Xinwen Zhang, and Xingyu Zheng for their valuable support and insightful suggestions on improving the manuscript.
\bibliographystyle{named}
\bibliography{ijcai25}

\clearpage
\twocolumn[%
  \begin{center}
    {\LARGE \bf Supplementary Material}
  \end{center}
  \vspace{1em}
]
\setcounter{section}{0}
\renewcommand{\thesection}{S\arabic{section}}  

\section{Dataset and Benchmark}
In this section, we give a description of the detail points in the Dataset section and Benchmark section in the body paper.

\subsection{Dataset}
\begin{figure}[h]
    \centering
    \includegraphics[scale=0.2]{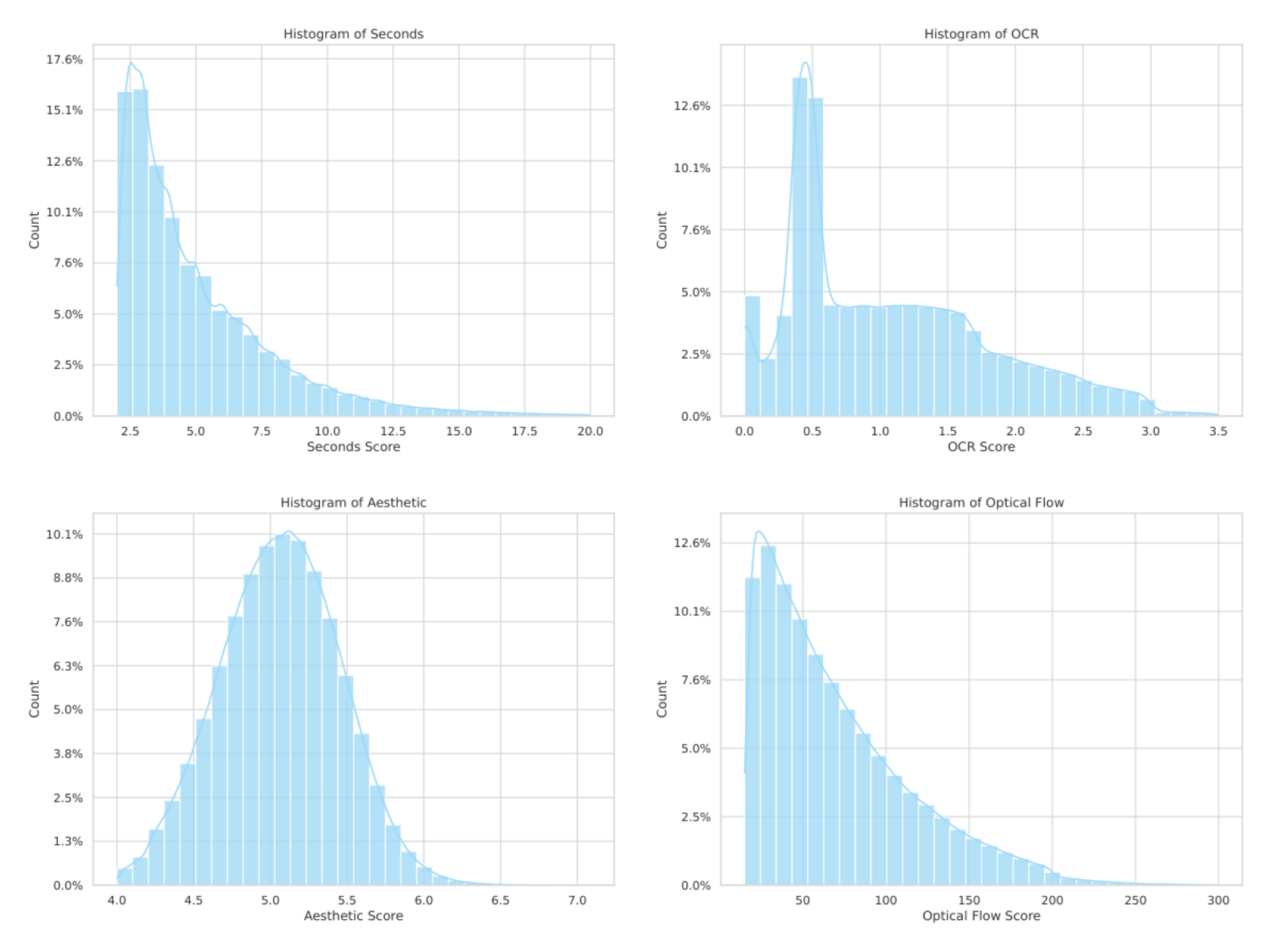}
    \caption{Distribution of Duration, OCR, Aesthetics, and Optical Flow in the Animation Dataset for Training}
    \label{fig:hist}
\end{figure}

First, we give the distribution for dimensions (text-cover region/OCR score, optical flow score, aesthetic score, and number of frames) of the animation dataset for model training. (shown in Fig. \ref{fig:hist})

As mentioned in the Animation Dataset Construction section on page 3, a few higher-quality clips will be finally filtered from the training set to further improve the model’s performance.
Compared with the initial screening, we use the fine-tuned filter model and set more stringent thresholds from these four dimensions to select higher-quality clips. 
The count of remaining high-quality clips is $0.5\%$ of the total.
Besides, we train several fine-grained data recognition models, e.g., talking, motion amplitude, and 2D/3D animation clip recognition. 
These models help us to control the distribution of the training clips, which can further improve the generation performance.

\subsection{Benchmark}
\noindent\textbf{Visual Motion}
When we train a motion model to evaluate the motion amplitude of the video clips (mentioned on page 5), the prompts in the inference step list as follows:

[\textit{'The protagonist has a large range of movement, such as running, jumping, dancing, or waving arms.',
'The protagonist remains stationary in the video with no apparent movement.'
}]

We can get a $1\times2$ similarity score from each input video clip and prompt. After the softmax function, we can get the final motion score.

\begin{figure}[h]
    \centering
    \includegraphics[scale=0.25]{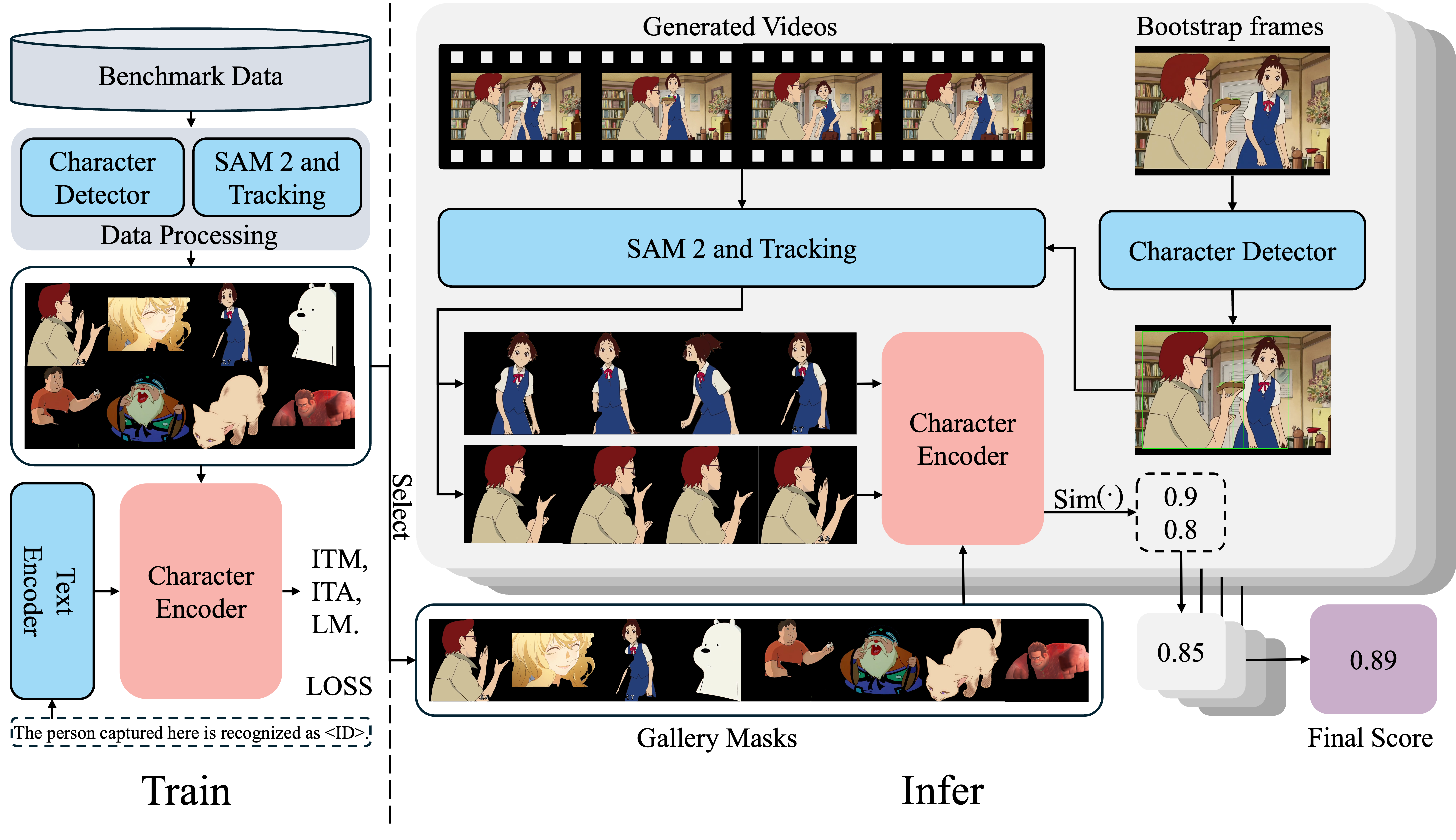}
    \caption{The Framework of Character Consistency: Training and Inference}
    \label{fig:ipconsis}
\end{figure}

\noindent\textbf{Character Consistency}
Fig. \ref{fig:ipconsis} gives the framework of character consistency model on page 6.
We leverage detection (GroudingDino), segmentation (SAM) and tracking tools to get the characters' masks on ground truth clips at first. Then, we finetune a character encoder module to extractor corresponding characters' features. 
In the infer step, we extract the characters' features from generation video clips, and get the final score through similarity calculation.


\begin{figure*}[h]
    \centering
    \includegraphics[scale=0.28]{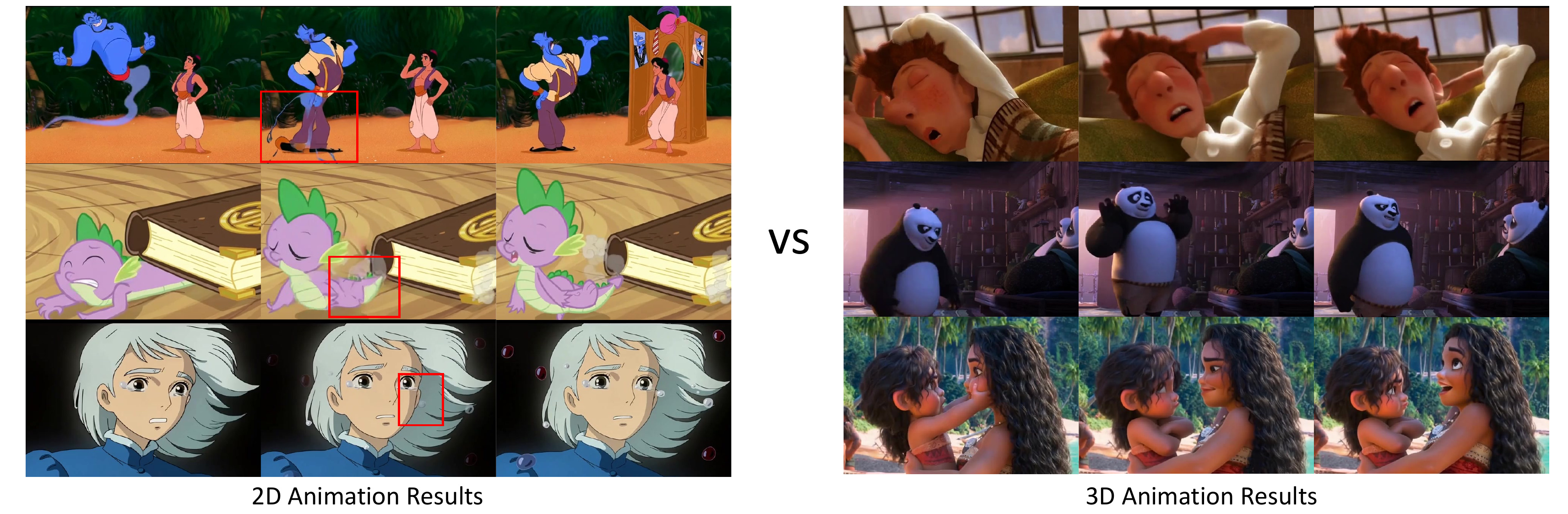}
    \caption{Comparision of 2D and 3D animation examples. The badcases in 2D animation are mainly due to exaggerated deformations, diverse appearances, and motions that violate physical laws.}
    \label{fig:2dvs3d}
\end{figure*}

\begin{figure*}[!h]
    \centering
    \includegraphics[width=0.9\textwidth]{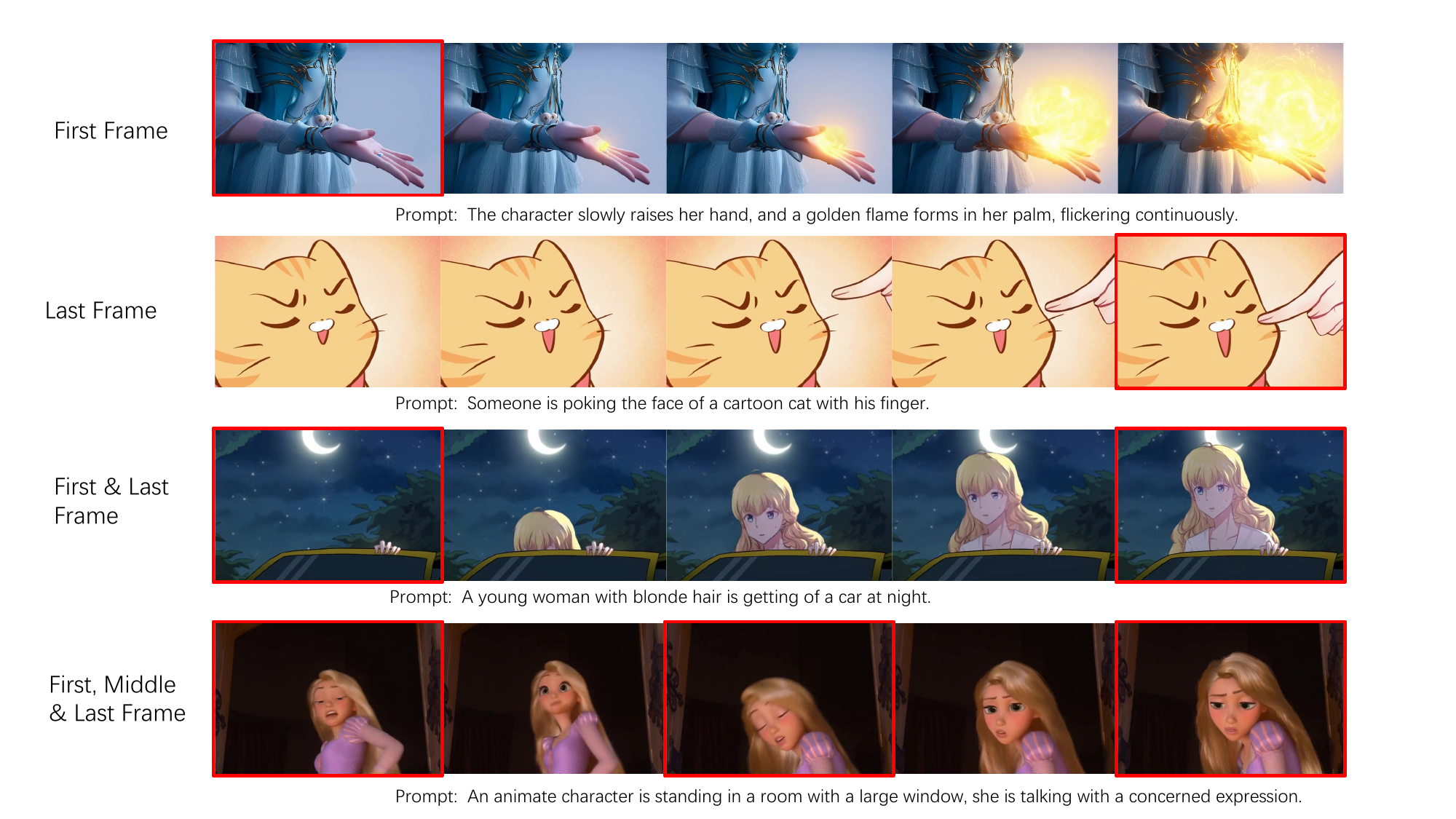}
    \caption{Illustration of different interpolation strategies. The images highlighted in red indicate the provided reference images. }
    \label{fig:frame_interpolation}
\end{figure*}

\begin{table*}[ht]
\centering
\caption{Quantitative Results on Vbench. (Note that AniSora-K denotes the results with keyframe interpolation, and AniSora-I denotes the interpolated average results of AniSora)}
\renewcommand{\arraystretch}{1.2} 
\setlength{\tabcolsep}{6pt}       
\scalebox{0.89}{
\begin{tabular}{ccccccccccc}
\hline
\multirow{3}{*}{\textbf{Method}} & \multicolumn{5}{c}{\textbf{Appearance}} &  \multicolumn{5}{c}{\textbf{Consistency}} \\
\cmidrule{2-5} \cmidrule{6-11}
& Motion & & Aesthetic  & Imaging & & I2V  & I2V & & Overall & Subject   \\
& Smoothness & & Quality & Quality & & Subject & Background & & Consistency &  Consistency   \\
\midrule
Opensora-Plan(V1.3) & 99.13  & & 53.21 & 65.11 & & 93.53 & 94.71 & & 21.67 & 88.86 \\
Opensora(V1.2) & 98.78  & & 54.30 & 68.44 & & 93.15 & 91.09 & & \textbf{22.68} & 87.71 \\
Vidu(V1.5) & 97.71  & & 53.68 & 69.23 & & 92.25 & 93.06 & & 20.87 & 88.27 \\
Cogvideo(5B-V1) & 97.67 & & \textbf{54.87} & 68.16 & & 90.68 & 91.79 & & 21.87 & 90.29 \\
MiniMax(I2V01) & 99.20 & & 54.66 & \textbf{71.67} & & 95.95 & \textbf{95.42} & & 21.82 & 93.62 \\
\midrule
AniSora & \textbf{99.34} & & 54.31 & 70.58 & & \textbf{97.52} & 95.04 & & 21.15 & \textbf{96.99} \\
AniSora-K & 99.12 & & 53.76 & 68.68 & & 95.13 & 93.36 & & 21.13 & 94.61 \\
AniSora-I & 99.31 & & 54.67 & 68.98 & & 94.16 & 92.38 & & 20.47 & 95.75 \\
GT & 98.72  & & 52.70 & 70.50 & & 96.02 & 95.03 & & 21.29 & 94.37 \\
\hline
\end{tabular}}
\label{tab:vbench result}
\end{table*}

\begin{figure*}[!h]
     \centering
     \includegraphics[scale=0.45]{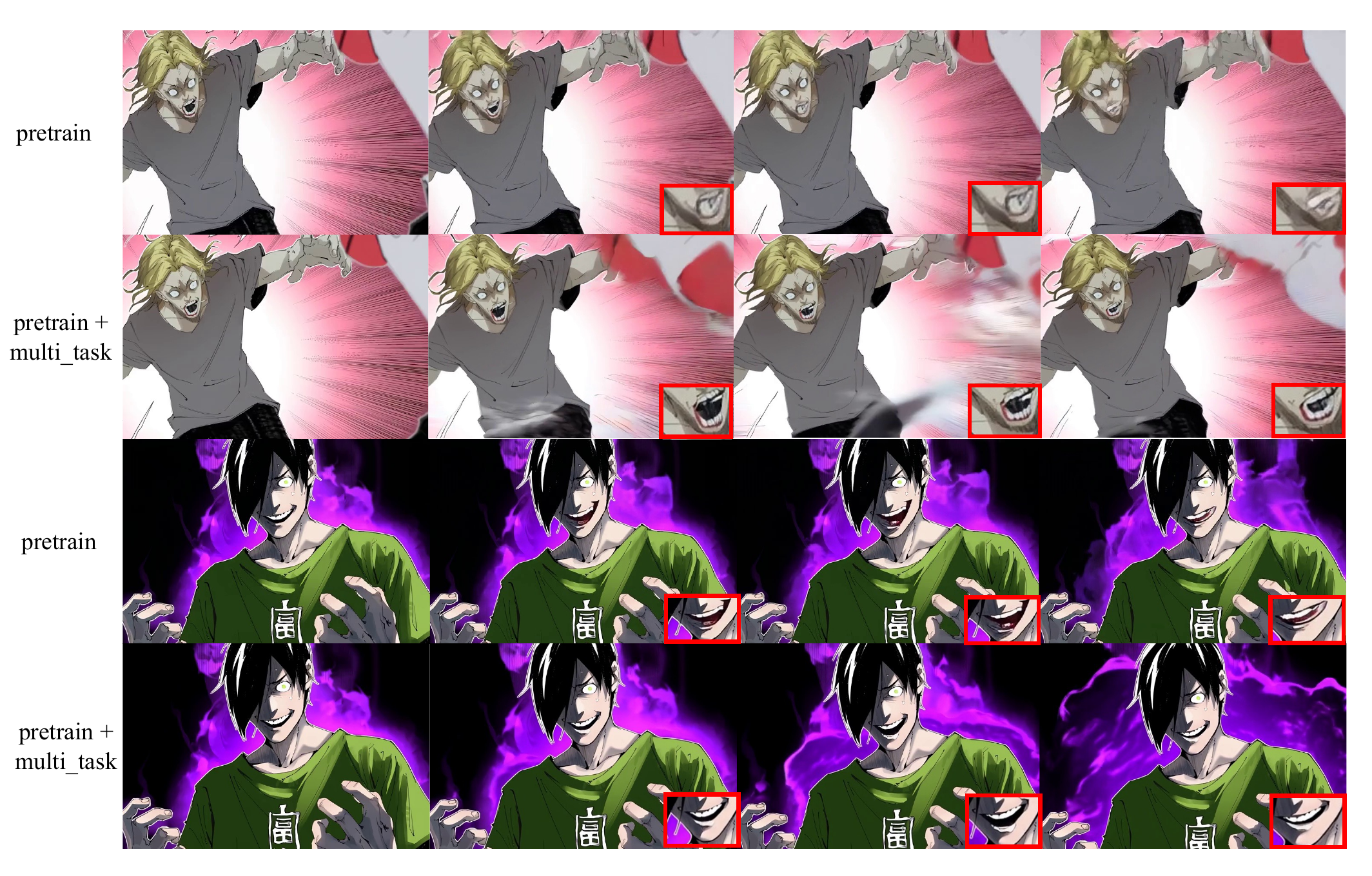}
     \caption{Comparision of results w/wo multi-task learning. The highlighted regions in red demonstrate significant improvements in stability and consistency after applying multi-task learning.}
     \label{fig:multi_task}
\end{figure*}

\section{Method}
\subsection{Supervised Fine-Tuning}

More effective training strategies are listed as follows:

\noindent\textbf{Weak to Strong} Our video generation model adopts a weak-to-strong training strategy to progressively enhance its learning capabilities across varying resolutions and frame rates. Initially, the model is trained on 480P videos at 8fps for 3 epochs, allowing it to capture basic spatiotemporal dynamics at a lower frame rate. Following this, the model undergoes training on 480P videos at 16fps for an additional 1.9 epochs, enabling it to refine its temporal consistency and adapt to higher frame rates. Finally, the model is fine-tuned on 720P videos at 16fps for 2.3 epochs, leveraging the previously learned features to generate high-resolution, temporally coherent video outputs. Additionally, we applied stricter filtering and produced a 1M ultra high-quality dataset for final-stage fine-tuning, significantly boosting high-resolution video quality.

\noindent\textbf{Removing Generated Subtitles} The presence of a significant number of videos with subtitles and platform watermarks in our training data led to the model occasionally generating such artifacts in its outputs. To mitigate this issue, we performed supervised fine-tuning using a curated dataset of videos entirely free of subtitles and watermarks. This dataset, consisting of 790k video clips, was constructed through proportional cropping of videos containing subtitles and the selection of clean, subtitle-free videos. Full-parameter fine-tuning was then applied to the model, and after 5.5k iterations, we observed that the model effectively eliminated the generation of subtitles and watermarks without compromising its overall performance.


\noindent\textbf{Temporal Multi-Resolution Training} Given the scarcity of high-quality animation data, we employ a mixed training strategy using video clips of varying durations to maximize data utilization. Specifically, a variable-length training approach is adopted, with training durations ranging from 2 to 8 seconds. This strategy enables our model to generate 720p video clips with flexible lengths between 2 and 8 seconds.

\begin{figure*}[!h]
    \centering
    \includegraphics[scale=0.4]{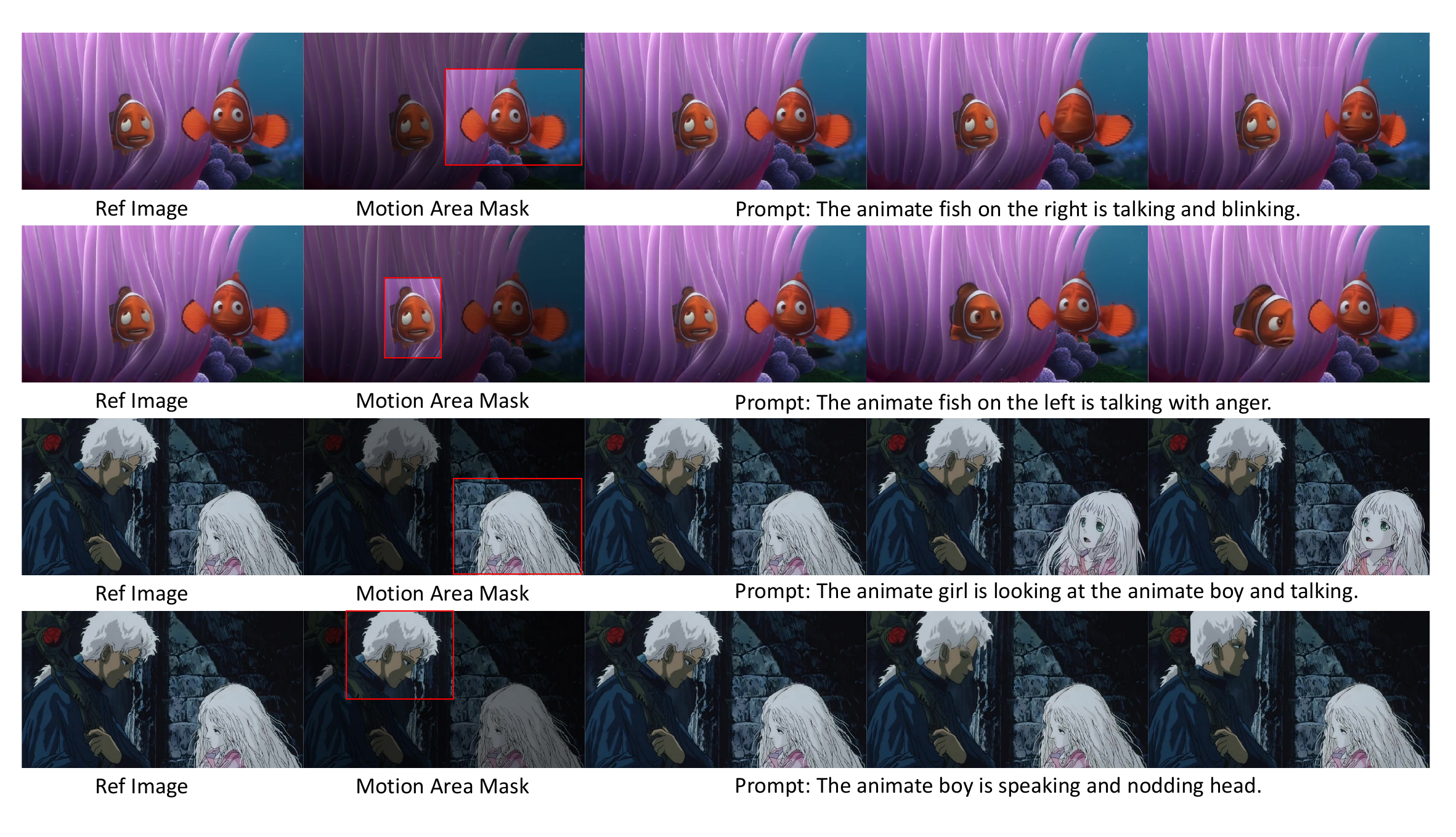}
    \caption{Examples of motion mask guidance. The first column shows the ref image, while the second column displays the mask. Animation creators can produce videos with fine-grained control over characters and backgrounds, ensuring alignment with various storylines.}
    \label{fig:motion_area}
\end{figure*}

\section{Experiment}

\subsection{Quantitative Results on Vbench}
Tab. \ref{tab:vbench result} gives the results on $7$ dimensions from Vbench.
It is easy to observe that the scores between different methods are a lack of differentiation.

\begin{table}[ht]
    \centering
    \caption{Benchmark Evaluation Results between 2D and 3D Generation Clips}
    \label{tab:benchmark_results_2d_3d}
    \begin{tabular}{lccc}
        \toprule
        Dims & 3D & 2D & All  \\     
        \midrule
        Visual Smooth & 69.44 & \textbf{71.69} & 71.47  \\
        Visual Motion & 45.59 & \textbf{70.29} & 47.94  \\
        Visual Appeal & 57.99 & \textbf{65.04} & 64.44  \\
        Text-Video Consistency & 72.86 & \textbf{72.93} & 72.92  \\
        Image-Video Consistency & 79.75 & \textbf{82.29} & 81.54 \\
        Character Consistency & 93.5 & \textbf{94.64} & 94.54  \\
        \bottomrule
    \end{tabular}
\end{table}

\subsection{2D and 3D Animation}
The Fig. \ref{fig:2dvs3d} demonstrates some results of 2D and 3D animation generation. As we discussed in the Experiment section, despite having a larger number of 2D data, generating high-quality 2D animation videos remains challenging due to the complexity of animation. Artifacts are more prevalent in 2D generation results, such as exaggerated deformations, more diverse character appearances, and motions that break the physical rules. For instance, in the third row of 2D examples, tears appear to floating in the air, making it more difficult for the model to capture the dynamic details accurately. In contrast, the motions rendered by the physics-based engines in 3D animations enable the model to achieve more reasonable results.

Moreover, Tab. \ref{tab:benchmark_results_2d_3d} gives the benchmark results among 2D, 3D, and total clips. The results support the observation that 3D generation results perform better than those of 2D, especially on motion score.


\subsection{Spatiotemporal Mask Module}
\subsubsection{Keyframe Guide Examples}

As shown in Fig. \ref{fig:frame_interpolation}, our unified framework supports different interpolation settings, enabling these functions to meet the demands of professional animation production. We observe that more guiding frames contribute to a more stable character identity and more precise actions align with creators. Nonetheless, amateur creators can still obtain satisfactory results by using just the first or last frame.


\begin{figure}[!h]
     \centering
     \includegraphics[scale=0.20]{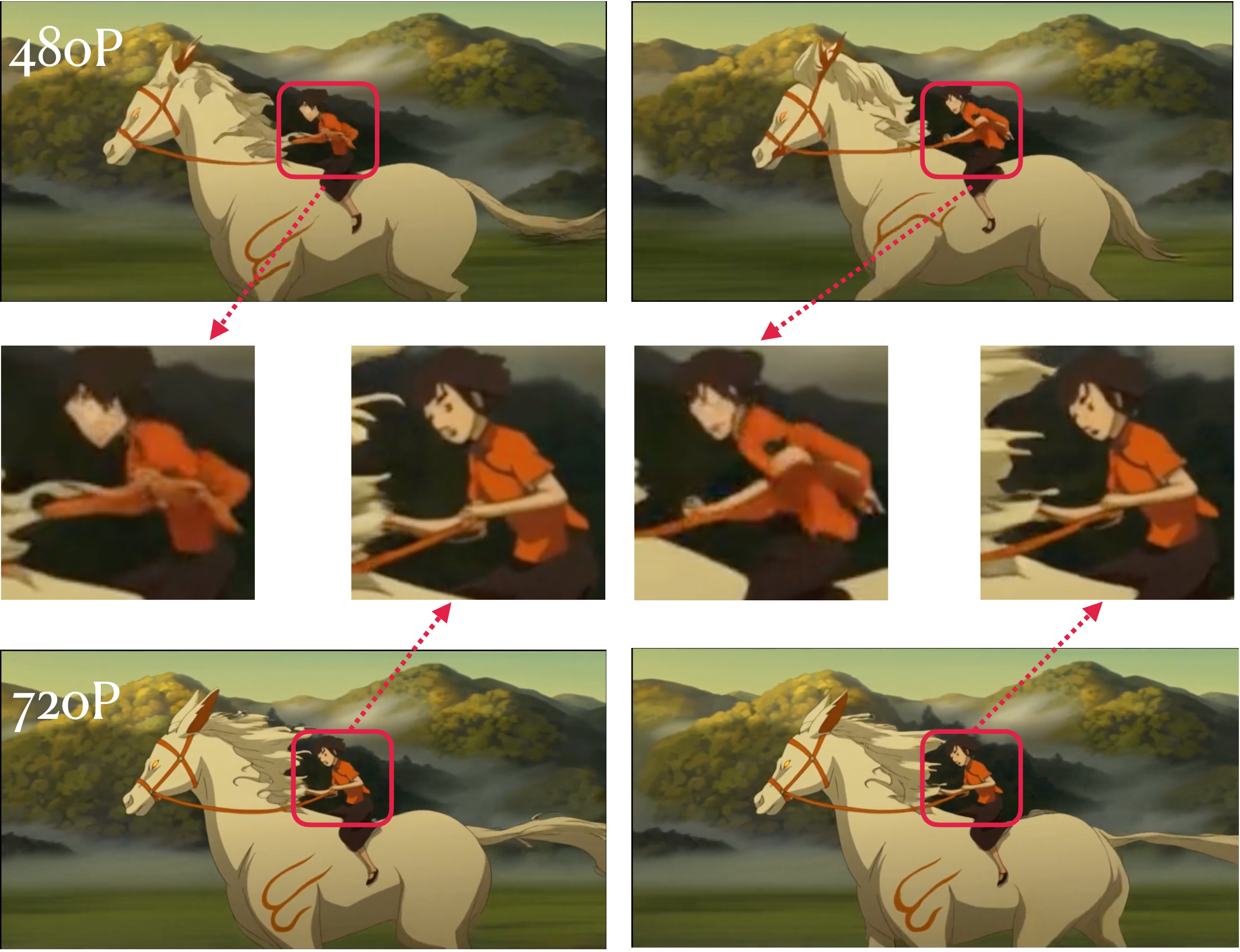}
     \caption{The figure compares the generation performance of 480P and 720P videos, highlighting that 720P achieves greater stability in the generation of details such as the character's facial features and hands.}
     \label{fig:720pvs480p}
\end{figure}

\subsubsection{Low-resolution vs High-resolution}
Fig. \ref{fig:720pvs480p} shows the comparison between 480P and 720P, where higher resolution improves temporal consistency and visual detail stability.

\subsubsection{Motion Area Condition}

Precise and effective control is especially crucial in the animation production pipeline. Fig. \ref{fig:motion_area} illustrates several motion mask guidance examples. 

\begin{figure*}[!h]
    \centering
    \includegraphics[width=\textwidth]{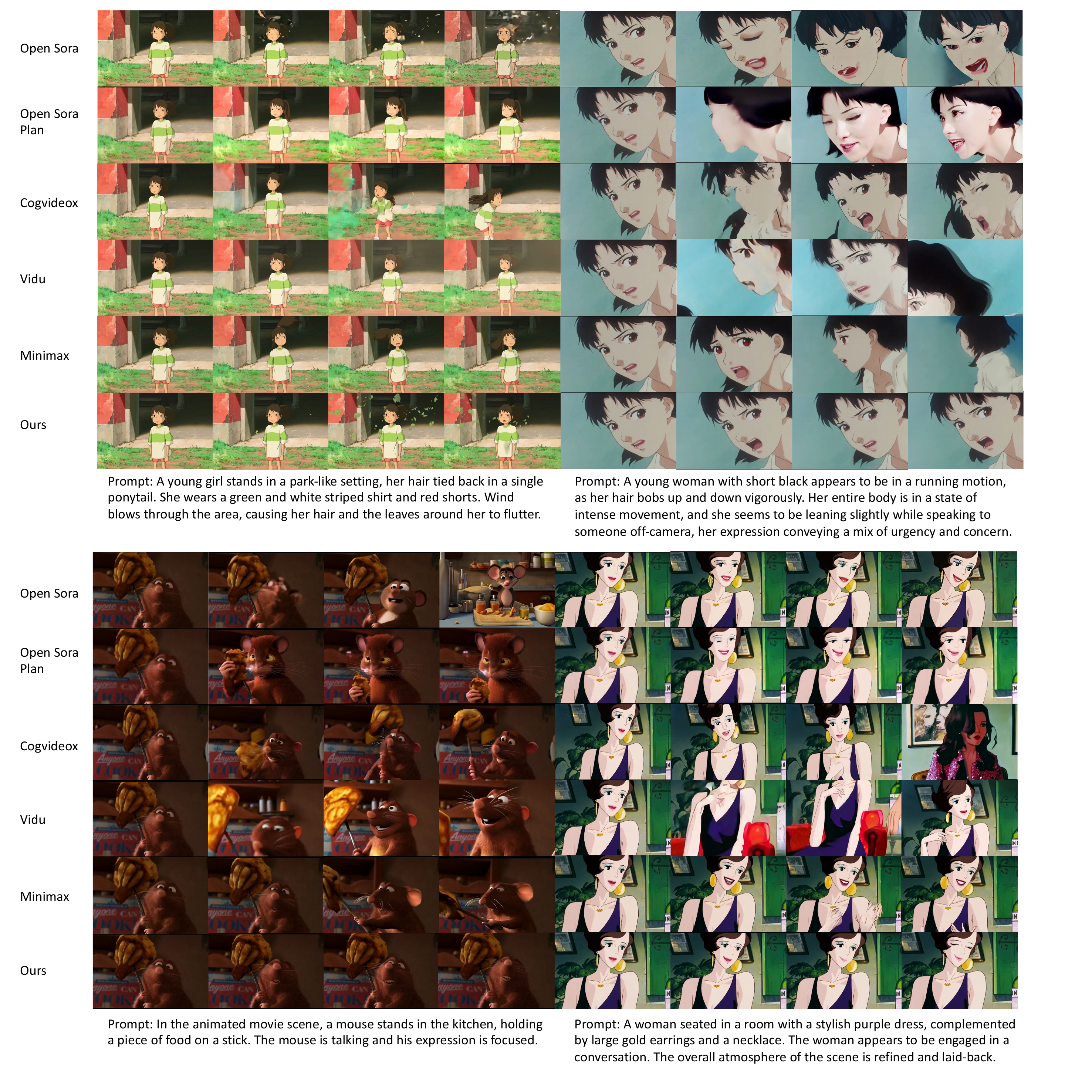}
    \caption{Comparison of our method with others using the first frame in the leftmost column as the guiding condition. Existing methods often struggle with animation data, leading to issues such as character identity shifts, unnatural dynamics, and motion blur.}
    \label{fig:method_comparision}
\end{figure*}

\subsubsection{Multi-Task Learning}
The diversity of anime styles presents a challenge for video generation. Although our model performs well in most styles, unique styles may result in inconsistencies, particularly in character details. To address this, we applied multi-task learning, combining image and video training to enhance the model’s adaptability to diverse styles. As shown in Fig. \ref{fig:multi_task}, without incorporating images, the model struggles to fully understand such styles, resulting in flaws in character detail generation. With the help of a small dataset of 270+ images, multi-task learning significantly improves the model’s ability to generate consistent and accurate details. With this technique, our unified framework can quickly adapt to more sophisticated styles, addressing the varied demands of creators.

High-resolution fine-tuning enhances the model's ability to capture intricate image details, as illustrated in Fig. \ref{fig:720pvs480p}. At a resolution of 480p, distortions in facial features can be observed. However, at 720p, the model not only maintains dynamic consistency but also preserves fine details in areas such as faces and hands.

\section{More Examples}
In this section, we show more generation examples compared with other methods, including Vidu, open-sora, open-sora plan, cogvideox, and minimax.
Our unified framework can effectively generate higher-quality animation videos that align more closely with the prompt while avoiding issues such as character distortion or motion blur. Examples are shown in Fig. \ref{fig:method_comparision}.



\end{document}